\documentclass[preprint]{aastex}

\newcommand{\lens}{PMN~J2004--1349}
\newcommand{\lensA}{PMN~J1838--3427}

\slugcomment{}

\begin{document}

\title{
A nearly symmetric double-image gravitational lens 
\footnote{ Based on observations using the Very Large Array (VLA) and
Very Long Baseline Array (VLBA); the Multi-Element Radio Linked
Interferometer Network (MERLIN); and the Blanco 4m telescope at the
Cerro Tololo Inter-American Observatory (CTIO). The VLA and VLBA are
operated by the National Radio Astronomy Observatory (NRAO), a
facility of the National Science Foundation (NSF) operated under
cooperative agreement by Associated Universities, Inc.\ MERLIN is a
UK national facility operated by the University of Manchester on
behalf of SERC. CTIO is operated by the Association of Universities
for Research in Astronomy Inc., under a cooperative agreement with the
NSF as part of the National Optical Astronomy Observatories. } }

\author{
Joshua N.\ Winn\altaffilmark{2,3},
Jacqueline N.\ Hewitt\altaffilmark{2},
Alok R.\ Patnaik\altaffilmark{4},
Paul L.\ Schechter\altaffilmark{2,3},
Robert A.\ Schommer\altaffilmark{5},
Sebastian Lopez\altaffilmark{6},
Jose Maza\altaffilmark{6},
Stefanie Wachter\altaffilmark{5}
}

\altaffiltext{2}{Department of Physics, Massachusetts Institute of Technology,
    Cambridge, MA 02139}
\altaffiltext{3}{Visiting Astronomer, Cerro Tololo Inter-American Observatory,
    National Optical Astronomy Observatories (NOAO)}
\altaffiltext{4}{Max-Planck-Institut f\"{u}r Radioastronomie,
    Auf dem H\"{u}gel 69, 53121 Bonn, Germany}
\altaffiltext{5}{Cerro Tololo Inter-American Observatory, National Optical
    Astronomy Observatories, Casilla 603, La Serena, Chile}
\altaffiltext{6}{Universidad de Chile, Santiago, Chile}

\begin{abstract}
We report the discovery of a new double-image gravitational lens
resulting from our search for lenses in the southern sky. Radio source
\lens~is composed of two compact components separated by $1\farcs13$
in VLA, MERLIN and VLBA images. The components have a flux ratio of
1:1 at radio frequencies ranging from 5~GHz to 22~GHz. The $I$-band
optical counterpart is also an equal double, with roughly the same
separation and position angle as the radio double. Upon subtraction of
the components from the $I$-band image, we identify a dim pattern of
residuals as the lens galaxy. While the present observations are
sufficient to establish that \lens~is a gravitational lens, additional
information will be necessary (such as the redshifts of the galaxy and
quasar, and precise astrometry and photometry of the lens galaxy)
before constructing detailed mass models.
\end{abstract}


\section{Introduction}
\label{sec:intro}
The new double-image gravitational lens presented in this paper was
found during a systematic search for lenses in the southern sky. Among
the reasons to search for new lenses are: to find objects well-suited
for determining the Hubble constant via time-delay measurements
\citep{refsdal,myers99b,schech00}; to constrain the cosmological
constant by measuring the incidence of multiple-image lensing
\citep{lambda0,lambda1,lambda2,lambda3}; and to study the mass
distributions (e.g., Keeton, Kochanek \& Seljak 1997; Rusin \& Ma
2000) and evolution \citep{castles1} of medium-redshift galaxies.

Our search methodology will be described in a future paper, but is
briefly as follows.  Our sample consists of southern radio sources
($0\arcdeg > \delta > -40\arcdeg$) with flat radio spectra ($\alpha
\geq -0.5$, where $S_{\nu} \propto \nu^{\alpha}$), as measured between
the 4.85~GHz Parkes-MIT-NRAO catalog \citep{pmn} and the 1.4~GHz NRAO
VLA Sky Survey \citep[NVSS]{nvss}.  This declination range is
relatively unexplored for lenses and therefore likely to contain
bright and useful specimens, and is far enough north to be within
reach of the VLA and VLBA, instruments that facilitate the search.

We obtained a snapshot of each object at 8.46~GHz with the VLA in its
A configuration.  Objects exhibiting multiple compact components
(about 5\% of the sample) were selected as lens candidates and
scheduled for appropriate follow-up radio and optical observations.
The goal of the follow-up observations is to determine whether the
components have comparable spectral properties and surface
brightnesses (as lensed images should have) and to search for a lens
galaxy.

This paper describes the second confirmed lens from our survey. The
first was \lensA, described by \citet{pmnj1838}.  Following the
organization of that paper, we end this introduction with a short
chronology of the discovery of \lens, then examine the evidence in
detail in subsequent sections.

The initial VLA image of 1998 May 18 contained two compact components
separated by $1\farcs13$. In the majority of such cases,
higher-resolution radio imaging reveals one component to be
flat-spectrum and compact, and the other to be steep-spectrum and
extended: a core-jet structure. However, both components were
unresolved in a MERLIN image obtained on 2000 April 1, which had 2.5
times the resolution of the VLA image. The components were also nearly
unresolved in a VLBA image obtained on 2000 April 28, which had 100
times the resolution of the VLA image.

With two compact components in such close proximity, \lens~was
established as either a binary quasar or gravitational
lens.\footnote{Here and elsewhere in this paper, we refer to the radio
components as ``quasars'' because they are compact radio sources with
pointlike optical counterparts, although strictly speaking they are
not spectroscopically verified quasars.}  The most conclusive way to
distinguish between these possibilities is to search for optical
evidence of a lens galaxy. Such evidence was found
in optical images obtained on 2000 July 25 with the Blanco 4m
telescope at CTIO.

Section~\ref{sec:radio} presents the radio properties of the system,
and \S~\ref{sec:opt} presents the optical properties.  Within the
latter section, \S~\ref{subsec:lensgalaxy} presents the evidence for
the lens galaxy, and \S~\ref{subsec:field} provides photometry of
\lens~and 5 nearby reference stars. In the last section we review the
evidence that \lens~is a gravitational lens and use the observed
properties to speculate on the nature of the lens galaxy and to
estimate the order of magnitude of the time delay.

\section{Radio images}
\label{sec:radio}

We present results from radio images obtained with the VLA, MERLIN and
VLBA, at frequencies ranging from 5~GHz to 22.4~GHz. In all the
images, there are two components with nearly equal fluxes, separated
by $1\farcs13$ at P.A.\ $60\fdg6$. We refer to the northeast component
as NE and the southwest component as SW. Based on the average of our
VLA measurements, the position of component NE is R.A.~(J2000)$~=
20^{{\mathrm h}}04^{{\mathrm m}}07^{{\mathrm s}}.091$, Dec.~(J2000)~$=
-13\arcdeg 49\arcmin 31\farcs07$, within $0\farcs1$.

\subsection{VLA images}
\label{subsec:vla}
We observed \lens~on 1998 May 18 for 35 seconds with the VLA at
8.46~GHz as part of the first phase of our lens search.  Radio source
3C286 was used to set the absolute flux density scale, following the
procedure suggested in the VLA Calibration Manual and adopting a flux
density of 5.52 Jy.

On 2000 November 1 we obtained additional VLA observations at
14.94~GHz and 22.46~GHz. These observations employed fast-switching
between the target and the phase calibration source J2011--157 (from
the VLA Calibration Manual). The total dwell time on \lens~was 5
minutes at 14.94~GHz and 8.3 minutes at 22.46~GHz. To set the absolute
flux density scale we observed J2355+498, a calibration source that is
monitored monthly by S.\ Myers and G.\ Taylor of the NRAO and has
proven to have a constant flux (within 5\%) at these frequencies. The
assumed flux densities of J2355+498 were 0.602~Jy at 14.94~GHz and
0.473~Jy at 22.46~GHz.

In all cases, the total observing bandwidth was 100~MHz. Calibration
was performed with AIPS using standard procedures, and imaging was
performed with Difmap \citep{difmap}. In all cases, two compact
components were obvious in the image prior to any deconvolution. We
fit a model consisting of two elliptical Gaussian components to the
visibility function using the ``modelfit'' utility of Difmap. After
repeated iterations of model-fitting and phase-only self-calibration,
we created a uniformly-weighted image with an elliptical Gaussian
restoring beam. The 8.46~GHz image is shown in Figure~\ref{fig:radio}.

The component fluxes are listed in Table~\ref{tbl:radio}, along with
the resolution and RMS level of the images. In every case, the
separation between NE and SW was consistent with the more precise VLBA
value listed in \S~\ref{subsec:vlba}. The Gaussian widths of NE and SW
converged to very low values.

\subsection{MERLIN image}
\label{subsec:merlin}
On 2000 April 1 we observed \lens~with MERLIN, using all 6 telescopes
of the array. The total observing bandwidth was 15 MHz centered at
4.994~GHz.  To calibrate the antenna phases, we switched between the
target and the nearby compact bright source J1939--1525 (from the VLBA
Calibrator Survey) with a cycle time of 5 minutes.  We divided the
observation into four segments, separated in time by about an hour, in
order to improve the $uv$-coverage. The total dwell time on \lens~was
59 minutes. Radio source 3C286 was used to set the absolute flux
density scale, assuming a flux density of 7.38 Jy on the shortest
baseline.  Calibration was performed at Jodrell Bank using standard
MERLIN software, and imaging was performed with Difmap.

Using the same procedure as was applied to the VLA data, we fit a
two-component model to the visibility function. After repeated
iterations of model fitting and phase-only self-calibration with a
5-minute solution interval, we arrived at our final image
(Figure~\ref{fig:radio}) and model (Table~\ref{tbl:radio}). The image
is uniformly weighted and restored with an elliptical Gaussian beam.
Both components were unresolved.

The gravitational deflection of light is achromatic in the regime of
lensing by galaxies, so the magnification ratio of point components
should be independent of frequency. The near-equality of flux density
ratios from 5~GHz to 22~GHz is therefore evidence that the two
components are gravitationally lensed images.  It is unlikely that two
unrelated radio sources, or even the members of a binary quasar, would
have identical continuum radio spectra. Stronger evidence
for lensing is presented in \S~\ref{subsec:lensgalaxy}.

We note that the flux density ratio measured at 8.5~GHz is not in
formal agreement with the others, although it is within $2\sigma$.
This discrepancy is not a serious challenge to the gravitational
lensing interpretation. The 8.5~GHz measurement was made 2 years
earlier than the others. Variability of the background source causes
the instantaneous flux density ratio to vary, because each lensed
image displays the source at a different moment in its history.

\subsection{VLBA image}
\label{subsec:vlba}
We examined the system with all ten antennas of the VLBA on 2000 April
28. The phase center was approximately the midpoint between NE and SW.
The observing bandwidth was 32 MHz centered at 4.975~GHz. This
bandwidth was divided into 4 intermediate frequencies, each of which
was subdivided into sixteen 0.5~MHz channels.  To calibrate the
antenna phases, delays, and fringe rates, we switched between
\lens~and J1939--1525 with a cycle time of 5 minutes. The total dwell
time on \lens~was 66 minutes.

Calibration, including fringe-fritting, was performed with standard
AIPS procedures. We used a fringe-fit interval of 2 minutes on the
phase reference source and applied interpolated solutions for delay
and rate to \lens. Prior to imaging, we reduced the data volume by
averaging in time into 6-second bins and in frequency into 1~MHz
bins. This level of sampling was just enough to prevent significant
bandwidth or time-average smearing over the required field of view.

\begin{figure}
\figurenum{1}
\plotone{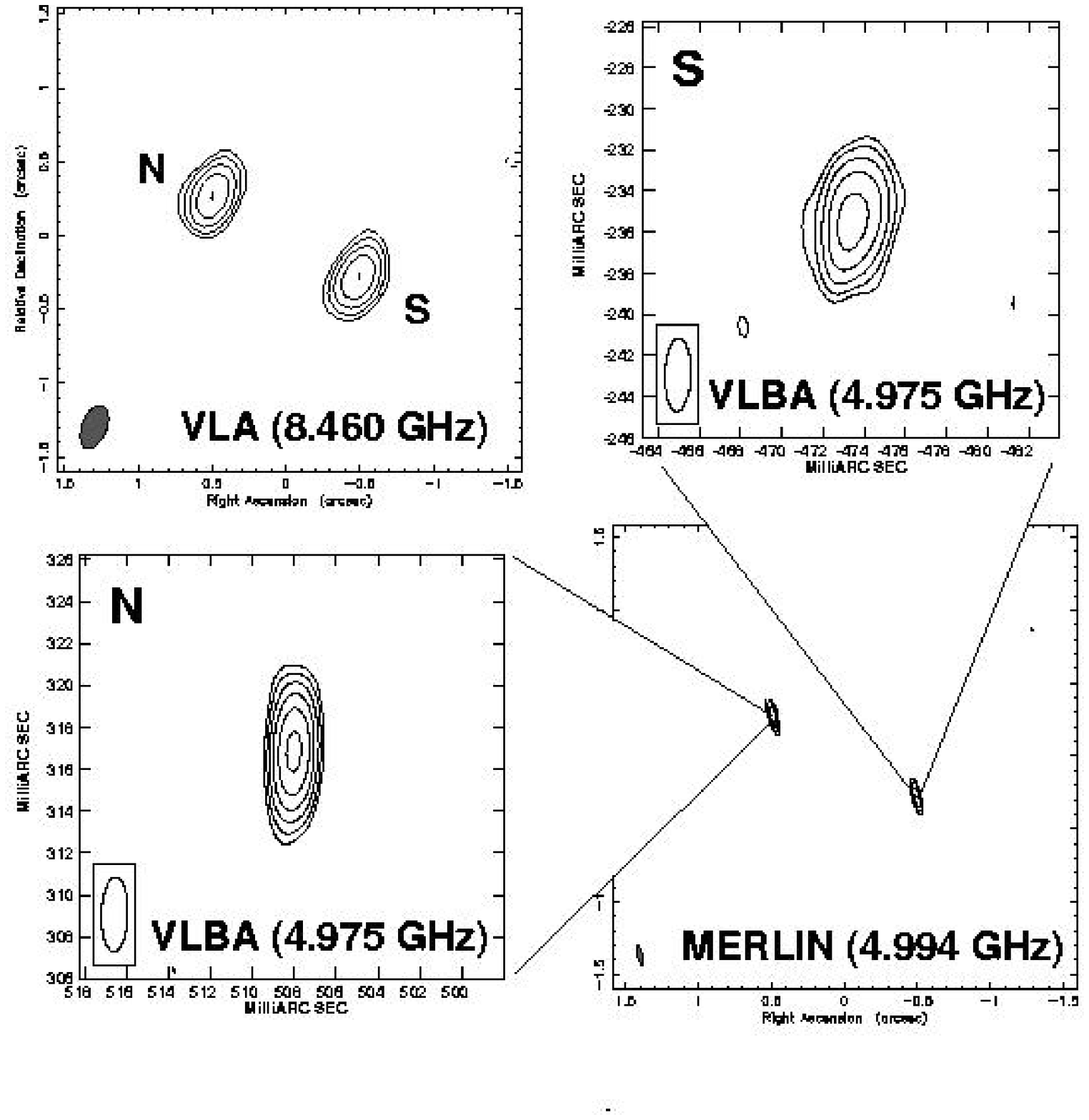}
\caption{ Radio images of \lens. In each panel, contours begin at
$3\sigma$ and increase by powers of 2, and the restoring beam is inset
in the lower left of the image.  The values of $\sigma$ and the beam
dimensions are in Table~\ref{tbl:radio}.  {\em Upper left.} VLA image
(3\arcsec $\times$ 3\arcsec).  {\em Lower right.} MERLIN image
(3\arcsec $\times$ 3\arcsec).  {\em Lower left.}  VLBA image (20 mas
$\times$ 20 mas) of component NE.  {\em Upper right.} VLBA image (20
mas $\times$ 20 mas) of component SW.  }
\label{fig:radio}
\end{figure}

For imaging, we employed the multiple-field implementation of the
Clean algorithm in AIPS, with a $512\times 512$ field (with 20 mas
pixels) centered on each component of \lens. The model constructed by
the Clean algorithm was then used to self-calibrate the antenna
phases with a solution interval of 2 minutes. This process, Cleaning
and phase-only self-calibration, was iterated 10 times before arriving
at the final model. The images based on this model, with uniform
weighting and an elliptical Gaussian restoring beam, are displayed in
Figure~\ref{fig:radio}. The restoring beam is elongated in the
north-south direction, with FWHM diameters $1.5\times 3.5$ mas.

Using the UVFIT task of AIPS, we fit a surface-brightness model
consisting of two elliptical Gaussian components to the visibility
data. To estimate of the uncertainty in each parameter, we observed
the scatter in the results of separately fitting the model to
individual 1~MHz spectral channels of data. The flux densities of the
best-fit model components are listed in Table~\ref{tbl:radio}. Again,
the flux density ratio is consistent with unity. The best-fit values
of $\Delta$R.A.\ and $\Delta$Decl.\ are printed in the table
caption. The models based on VLA and MERLIN data have separations that
are consistent with these values, which are the most precise.

The widths of component NE converged to values smaller than half the
width of the restoring beam. We conclude NE is unresolved. Component
SW converged to $2.06\times 0.76$ mas at a position angle of
$-29\arcdeg$, with a scatter of 0.5 mas in each width and $16\arcdeg$
in angle. Because the major axis is larger than the restoring beam, it
appears that SW is resolved along one direction. These results are
consistent with the visual appearance of the images in
Figure~\ref{fig:radio}.

If component SW is truly resolved, then it has a lower surface
brightness than NE, which complicates the hypothesis that \lens~is a
gravitational lens.  Gravitational deflection conserves surface
brightness. Lensed images are therefore expected to have the same
surface brightness in the absence of complications, such as
propagation effects that are different for each image path from source
to observer.

The width of SW is not due to time-average or bandwidth smearing,
which would in any case affect both components equally. Neither can
the width be attributed solely to phase decorrelation, because the
width does not vary significantly as the self-calibration solution
interval is altered, and because phase decorrelation would also affect
both components. We conclude that SW is indeed resolved.  This
conclusion is worth checking with further VLBI observations, because
the width is comparable to the resolution of the present data, and the
$uv$-coverage of this short observation was poor. This point is
discussed further in \S~\ref{subsec:summary}.

\subsection{Total flux density}

\begin{figure}
\figurenum{2}
\plotone{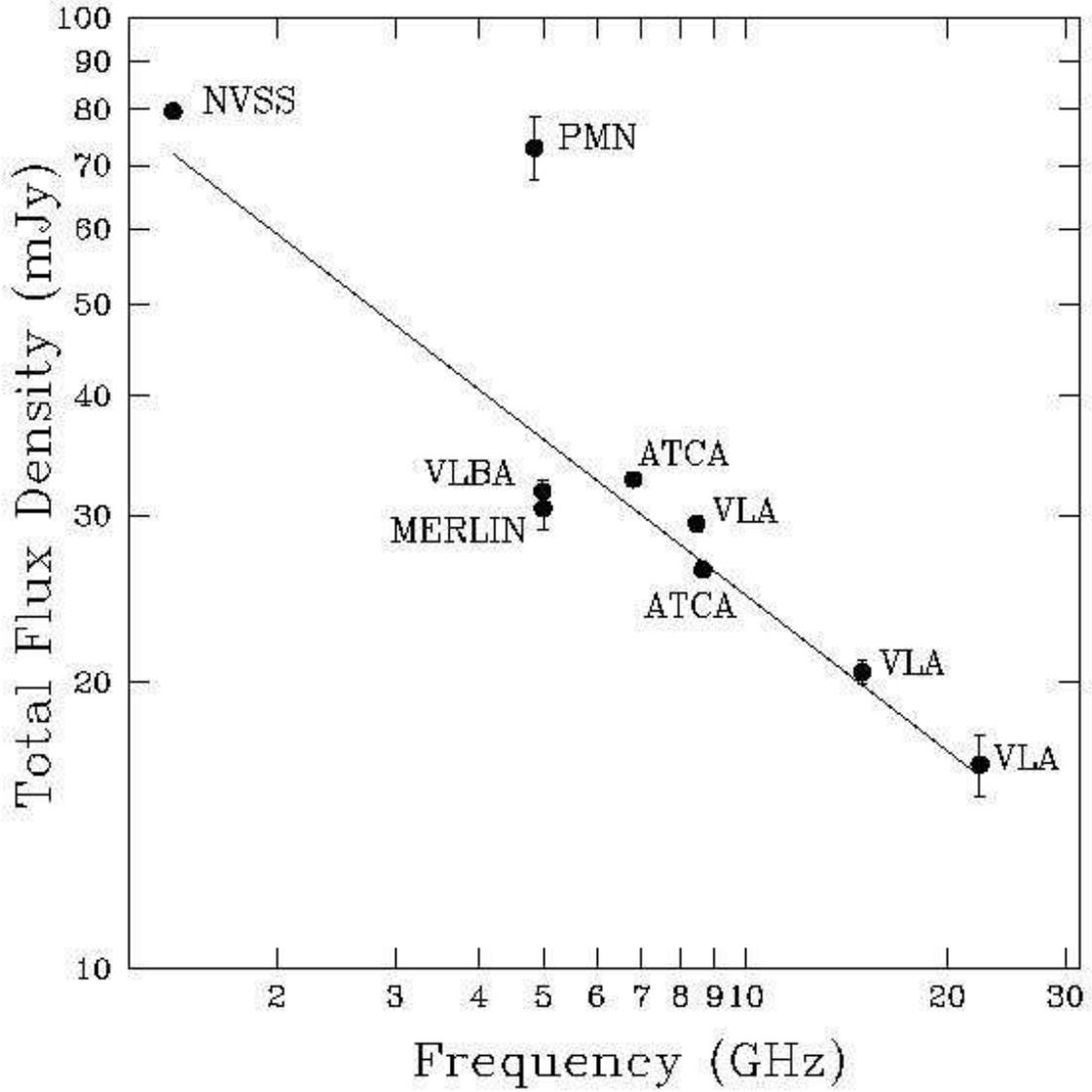}
\caption{ Total flux density of \lens~as a function of radio
frequency, on a log-log scale. The points labeled VLBA, MERLIN, ATCA
and VLA are from measurements described in this paper.  The points
labeled PMN and NVSS are drawn from those radio catalogs.  The solid
line is the best fit after disregarding the PMN point, and has slope
$\alpha = -0.54$. Where error bars are not shown, they are
comparable or smaller than the symbol size.}
\label{fig:totalflux}
\end{figure}

Assessing the variability of gravitational lenses is important because
flux monitoring of variable systems offers potentially significant
scientific rewards, including the Hubble constant
\citep{refsdal,myers99b,schech00} and the characterization of compact
masses in the lens galaxy \citep{koopmans}.
Figure~\ref{fig:totalflux} is a logarithmic plot of the total flux
density of \lens~as a function of radio frequency. It includes the
entries from the PMN \citep{pmnt} and NVSS \citep{nvss} catalogs; the
measurements with the VLA, VLBA and MERLIN described in this paper;
and also two measurements with the Australia Telescope Compact Array
(ATCA) that were taken on 2000 September 26 by J.E.J.\ Lovell (private
communication).

With the exception of the PMN flux density, the spectrum is a power
law with a best-fit spectral index of $\alpha = -0.54\pm 0.10$ (where
$S_\nu \sim \nu^{\alpha}$). This is a typical value for radio-loud
quasars. The large discrepancy between the PMN and the power law
spectrum has two possible explanations. First, it could be the result
of variability. The PMN flux is from 1990 whereas the others are from
2000. Second, the PMN flux is based on a single-dish measurement,
whereas the others are interferometric. Emission that is smooth on
scales larger than several arcseconds would be detected in the PMN
survey but invisible to the interferometers.

Variability is the more likely explanation. There is independent
evidence that \lens~is variable on a time scale of years, from the
discrepancy between the VLA (1998) and ATCA (2000) flux densities at
8.5~GHz. There is no independent evidence for extended emission nor is
such emission expected from typical radio quasars. The NVSS survey
used a $45\arcsec$ beam and might have been expected to detect this
diffuse flux, but instead the NVSS flux lies along the same power-law
line as the other interferometric measurements. There are no other
sources in the NVSS catalog within $5\arcmin$ of \lens.

\section{Optical images}
\label{sec:opt}

On 2000 July 25 we obtained optical images of \lens~with the Blanco 4m
telescope at CTIO. We used the Mosaic II CCD, which has eight
$2048\times 4096$ CCDs that are individually amplified and read out
in pairs. We centered the target in chip \#2.  The night was
photometric.  Table~\S~\ref{tbl:journal} reports the filter, exposure
time, airmass, and seeing of each observation.

The images from chip \#2 were extracted and corrected for cross-talk
from the paired amplifier.  These images were then bias-subtracted and
flat-fielded with standard IRAF\footnote{ IRAF is distributed by the
National Optical Astronomy Observatories, which are operated by the
Association of Universities for Research in Astronomy, Inc., under
cooperative agreement with the National Science Foundation.  }
procedures. In addition, for each of the two $I$-band images, sky
fringes were removed by subtracting an appropriately-scaled fringe
template for chip \#2, which was kindly supplied by R.C.\
Dohm-Palmer. The two frames were then registered and added to create
our final $I$-band image.  The next section describes the optical
counterpart of \lens~and, in particular, evidence of a lens galaxy in
the $I$-band image. Section~\ref{subsec:field} provides photometry of
\lens~and 5 nearby stars.

\subsection{Evidence of a lens galaxy}
\label{subsec:lensgalaxy}
\begin{figure}
\figurenum{3}
\plotone{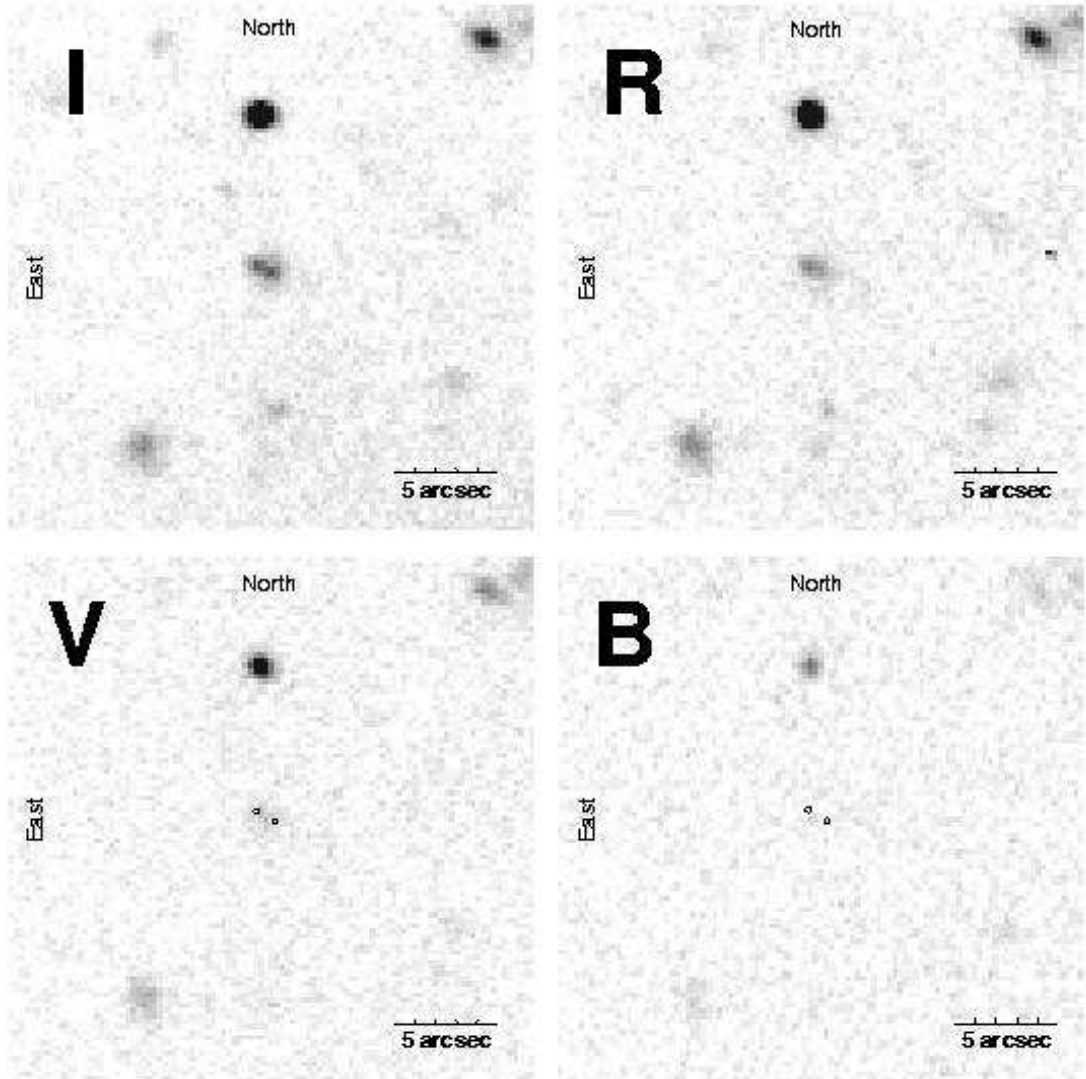}
\caption{ Optical counterpart of \lens~as revealed by $BVRI$ images
from CTIO (see \S~\ref{sec:opt}.  Each panel is a $27\arcsec\times
27\arcsec$ subraster with a logarithmic grayscale. North is up and
east is left.  In the $V$ and $B$ panels, the expected quasar
locations are marked with circles.  }
\label{fig:opt}
\end{figure}

The four panels of Figure~\ref{fig:opt} display the optical
counterpart as viewed through each of the four filters.  In $I$-band
the quasars are easily distinguished, and it is apparent they have
roughly the same separation, position angle and flux ratio as the
radio double. In $R$ the object is not as obviously double, and is not
as symmetrical. The $V$-band counterpart is a single faint source
lying between the expected quasar positions, which are marked with
circles. In $B$ the counterpart is barely detected if at all.

Were \lens~a binary quasar, the optical counterpart would be double in
all filters (with sufficient resolution and signal-to-noise ratio),
and the separation would match the radio separation within the
observational uncertainties.  Our aim in this section is to
demonstrate this is not the case.  Rather, the actual morphology and
its filter-dependence provides evidence of another light source
between the quasars, which we believe is the lens galaxy.

\begin{figure}
\figurenum{4}
\plotone{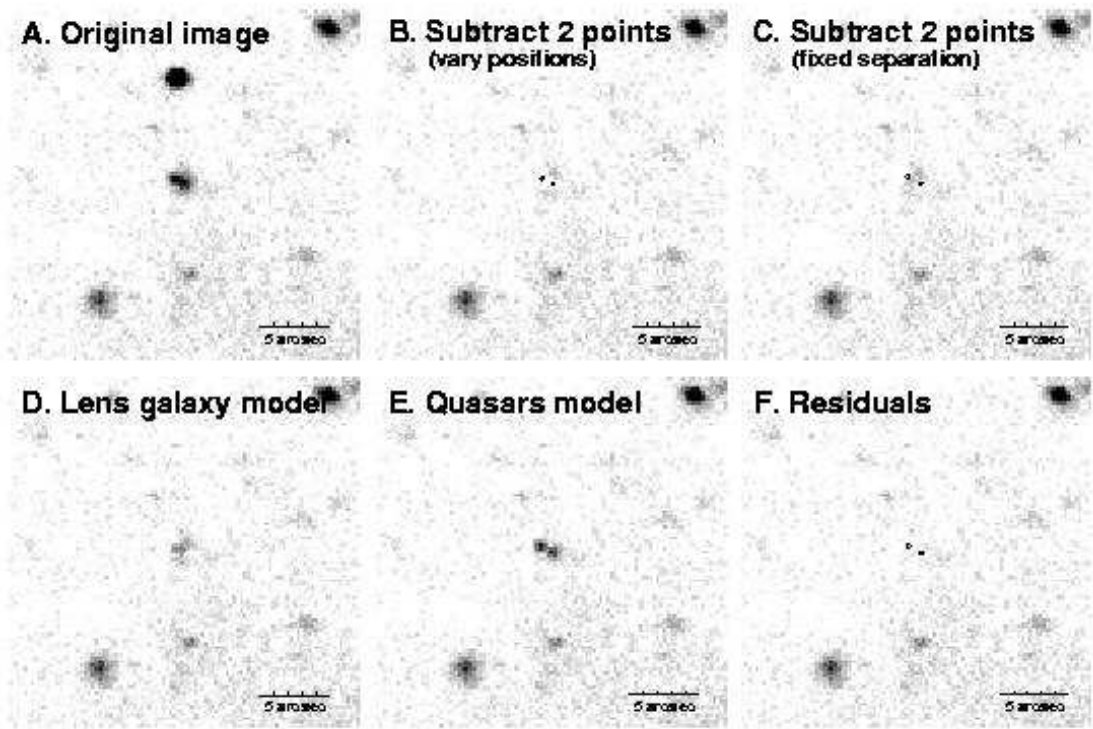}
\caption{ Photometric model-fitting to the $I$-band image of \lens.
Each panel is a $27\arcsec\times 27\arcsec$ subraster with a
logarithmic grayscale. North is up and east is left.  {\bf A.}
Original $I$-band image.  {\bf B.} Residual image after subtraction of
a 2-component model in which the positions and magnitudes of the
components were allowed to vary.  The locations of the subtracted
components are marked.  {\bf C.} Residual image after subtraction of a
2-component model in which the relative separation was fixed at the
VLBA value.  The locations of the subtracted components are marked.
{\bf D.} Residual image after subtraction of components NE and SW of
the final 5-component photometric model.  {\bf E.} Residual image
after subtraction of components G1--3 of the final 5-component
photometric model.  {\bf F.} Residual image after subtraction of the
entire 5-component photometric model. The positions of NE and SW are
marked.  }
\label{fig:resid}
\end{figure}

The most important evidence is in the $I$-band image, which is
reproduced at higher contrast in Panel A of
Figure~\ref{fig:resid}. With the DAOPHOT package in IRAF, we
constructed an empirical PSF of radius $6\farcs7$, using the
signal-weighted average of several bright and isolated field stars. We
then fit a model consisting of two point sources to the optical
counterpart of \lens. The best-fit separation was $880\pm 30$ mas
along a position angle of $62\arcdeg$, and the NE/SW flux ratio was
$1.01\pm 0.11$. Compared to the radio double ($1126.2\pm 0.2$ mas,
P.A.\ $61\arcdeg$) the optical double is along the same axis but is
significantly smaller.

Panel B is the residual image after the 2-point model has been
subtracted. The positions of the subtracted components are marked with
small circles. There are significant (4$\sigma$) positive residuals
just northwest and southeast of the axis of the double.  A natural way
to explain both the smaller optical separation and the pattern of
residuals is that there is a diffuse source of unmodeled light
centered between the quasars. The model components have been drawn
together to account for this extra flux.

To ensure that the diffuse flux is not an artifact of imperfect PSF
fitting, we created an artificial double with the same position angle
and separation as the radio double, using the image of an isolated
star in the $I$-band image that is comparable in brightness to
\lens. After applying the same PSF-fitting procedure as above, there
was no trace of residual flux at the level seen in the actual data.

Next, convinced of the reality of the diffuse flux between the
quasars, we fit a two-point model to the original image in which the
relative separation between the points was fixed at the VLBA value.
The NE/SW flux ratio converged to $1.00\pm 0.11$. Panel C of
Figure~\ref{fig:resid} shows the residual image, in which the diffuse
flux is seen to extend continuously across the axis joining the
quasars. This diffuse flux is naturally interpreted as a lens galaxy.

Based on the $V$ and $R$ morphology, the galaxy appears to be bluer
than the quasars. This explains why there is a light source between
the expected quasar positions in the $V$-band image, but the quasars
themselves are not obvious. It may also explain why the $R$-band
components are not as clearly separate as in $I$, despite a comparable
resolution and signal-to-noise ratio. This point is discussed
further in \S~\ref{subsec:nature}.

\subsection{Photometry of \lens~and field stars}
\label{subsec:field}

To represent the galaxy, we added a third point (G1) to our model
located exactly halfway between the quasars, and simultaneously solved
for all three fluxes. There were still significant residuals to the
north and south of G1, so we added two additional points (G2 and G3)
located at the peak residuals. Our final photometric model therefore
contained two points representing the quasars (NE and SW), and three
points for the galaxy (G1--3).  The relative separations were fixed at
the values printed in Table~\ref{tbl:photomodel}, and the fluxes were
allowed to vary simultaneously.  Panel D of Figure~\ref{fig:resid}
shows the $I$-band image with the quasars subtracted, in order to
display the lens galaxy model. In Panel E, the galaxy model has been
subtracted, highlighting the quasars. Finally, Panel F shows the
residual image after all five components have been subtracted.  We
applied the same 5-component model to the $R$- and $V$-band images.

\begin{figure}
\figurenum{5}
\plotone{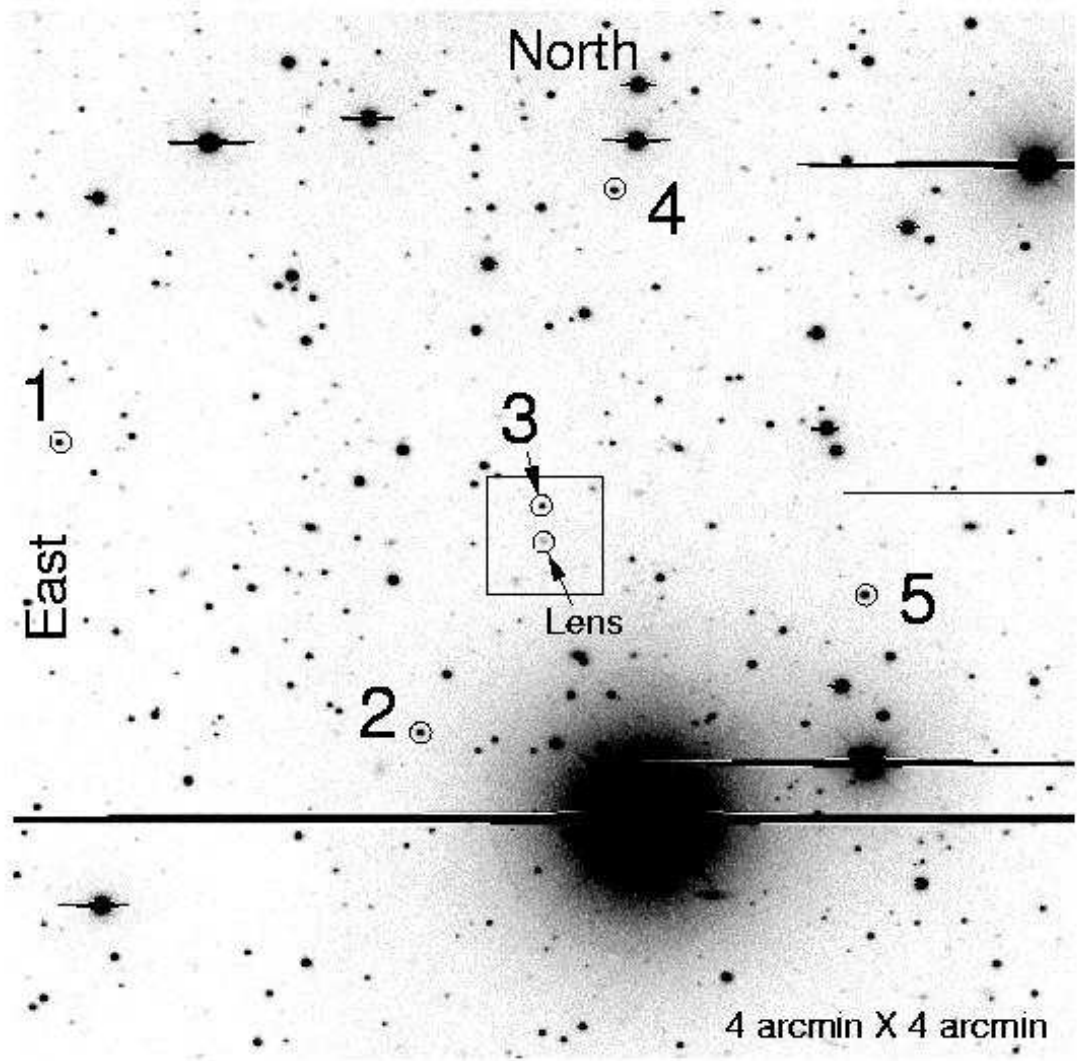}
\caption{ Wide-field ($4\arcmin\times 4\arcmin$) $I$-band image
centered on \lens. The 5 reference stars discussed in
\S~\ref{subsec:field} are circled and numbered.  Magnitudes and
positions of these stars and of \lens~are printed in
Table~\ref{tbl:photometry}.  The square box shows the field of view of
the images of Figures~\ref{fig:opt}~and~\ref{fig:resid}.  }
\label{fig:field}
\end{figure}

To establish a local photometric reference system for future
observations, we used the same PSF template to find the instrumental
magnitudes of 5 field stars. These stars are circled and labeled in
Figure~\ref{fig:field}. For photometric calibration we observed the
SA110 field described by \citet{landolt}. Star \#361 was used to
determine the magnitude zero point, using an aperture diameter of
$14\arcsec$. We adopted ``typical'' CTIO extinction coefficients of
$k_I = 0.06$, $k_R = 0.11$, $k_V = 0.15$, and $k_B = 0.28$
\citep{landolt}.

The calibrated magnitudes of the reference stars and the components of
\lens~are listed in Table~\ref{tbl:photometry}.  In $V$, the flux of
component SE converged to zero, so we report only a lower limit on the
magnitude. The non-detection in $B$ implies $B>24.3$ for all
components.

It is important to note that the quoted uncertainties are the
statistical errors in the fitting procedure, and therefore apply only
to magnitude differences. The calibration introduces an overall error
of 0.05 magnitudes that affects all entries identically. Furthermore,
the quoted uncertainties for the components of \lens~may be
underestimated because they are internal to our choice of photometric
model. Because all the components lie within two seeing discs, the
magnitudes are covariant.

\section{Discussion}
\label{sec:discussion}

\subsection{Summary of evidence for lensing}
\label{subsec:summary}

We now summarize the argument that \lens~is a gravitational lens. It
consists of two radio components, each of which is compact on
milliarcsecond scales and has a pointlike optical counterpart. This
implies \lens~is a pair of radio-loud quasars, and invites three
possible interpretations: a chance alignment of unrelated quasars, a
binary quasar, or a pair of lensed images of a single quasar.

The small separation ($1\farcs13$) argues against the chance alignment
and binary quasar hypotheses. Using the analysis of faint 5~GHz radio
source counts by \citet{langston}, the probability of such a close
alignment of two unrelated sources brighter than 15~mJy is less than
$2\times10^{-6}$. The chance of finding such a pair among our initial
sample of $\sim 4000$ radio sources is therefore less than $1\%$.
Furthermore, the small separation is more typical of lensed quasars
($0\farcs5 - 2\arcsec$) than binary quasars ($> 3\arcsec$; see e.g.\
\citet{binqso1,binqso2}).

The observation that the quasars have nearly the same flux ratio
($\sim$ 1:1) at 5~GHz, 8.5~GHz, 15~GHz and 22~GHz would require a
coincidence under the chance alignment and binary quasar hypotheses,
but is a natural consequence of the lensing hypothesis. The strongest
evidence for lensing is the presence of a diffuse source of light in
the $I$-band image centered between the quasar positions, which is
naturally interpreted as a lens galaxy.

We believe this argument is conclusive. However, the lensing
hypothesis is complicated by the observation that the quasars have
different angular sizes in our VLBA images
(\S~\ref{subsec:vlba}). Because gravitational lensing conserves
surface brightness, a pair of lensed images with equal flux should
have equal angular sizes. Our data suggest that the SW quasar is
resolved along one direction and the NE quasar is unresolved. If this
observation withstands further scrutiny, there must be a propagation
phenomenon that causes the surface brightnesses of the lensed images
to differ. One possibility is differential scatter-broadening of the
components by plasma in the lens galaxy or our own Galaxy. This
hypothesis can be tested by determining whether the angular size
varies characteristically as the square of the observing
wavelength. One might also determine whether the percentage
polarization or amount of Faraday rotation of the components are
different, which would be independent evidence of different electron
column densities.

\subsection{Nature of the lens galaxy}
\label{subsec:nature}

The photometry described in \S~\ref{subsec:field} was used to estimate
the magnitudes of the lens galaxy (see Table \ref{tbl:photometry}). As
described in that section, the quoted uncertainties are probably
underestimates, because the galaxy is a low-surface-brightness object
located between two quasars that are separated by only one seeing
disc. It is even possible that what we have called the ``lens galaxy''
is more than one galaxy. Despite these caveats, our estimates can be
used to speculate on the nature of the lens galaxy, pending
higher-resolution observations with adaptive optics or the {\em Hubble
Space Telescope}. As a first step, we correct for Galactic extinction
using the values of \citet{extinction}, obtaining $I=21.5$, $R=22.6$
and $V=23.2$ (within $\sim 0.2$ mag).

Are these magnitudes consistent with the magnitude of a galaxy
sufficiently massive to produce the observed image separation? The
image separation produced by a lens galaxy with velocity dispersion
$\sigma$ is roughly
\begin{equation}
\Delta\theta \sim \frac{4\pi\sigma^2}{c^2} = (1\farcs4)\left(\frac{\sigma}{ 220 \mbox{~km s}^{-1}} \right)^2 .
\end{equation}
The separation of $1\farcs13$ observed in \lens~therefore indicates a
galaxy that is not much less massive than a typical $L_*$ galaxy.  The
magnitudes and colors of typical $L_*$ spiral and elliptical galaxies,
over a range of redshifts, have been computed by \citet{lehar} using
several standard models for galaxy evolution.  According to these
calculations, an $I$-band magnitude of 21.5 implies a redshift between
0.5 and 1 (see Fig.\ 3 of Leh\'{a}r et al.\ 2000). This is the range
that is typical of lens galaxies, so the magnitude of the diffuse flux
is certainly consistent with the lensing hypothesis.

Furthermore, Lehar et al.\ (2000) show that for a given $I$-band
magnitude, the $V-I$ color is sometimes a useful diagnostic of galaxy
morphology. In particular, for the redshift range $0.5<z<1$,
early-type galaxies tend to be redder ($V-I \sim 3$) than late-type
galaxies ($V-I \sim 1.6$). In this sense our lens galaxy photometry
($V-I\sim 1.7\pm 0.3$) favors a late-type galaxy. This statement
depends upon the chosen model for the timing of star formation, and
might be false for an early-type galaxies with a very late
starburst. In addition, the conclusion may be sensitive to our choice
of photometric model for the lens galaxy. Nevertheless, our working
hypothesis is that the lens galaxy is a spiral galaxy with a redshift
between 0.5 and 1.

\subsection{Magnification ratio and time delay}
\label{subsec:timedelay}

In order to use a lensed quasar to determine the Hubble constant, one
must not only measure the time delay $\Delta t$ between the flux
variations of the lensed images, but also construct a model of the
gravitational potential of the lens galaxy that can be used to predict
the quantity $H_0 \Delta t$. The present observations do not allow for
detailed mass modeling. Deeper and higher-resolution images,
especially with adaptive optics or the {\em Hubble Space Telescope},
will be necessary for precise photometry and astrometry of the lens
galaxy. Optical spectroscopy will also be essential, in order to
obtain the redshifts of the background source and the lens. Despite
the faintness of the quasars and the lens galaxy, low-resolution
spectroscopy should not be a problem for an 8m-class telescope. In the
meantime we can use the few observed properties of \lens~to predict
the order of magnitude of the time delay.

The simplest plausible model for a lens galaxy is a singular
isothermal sphere, for which the time delay is
\begin{equation}
\Delta t = \frac{1}{2c} \left(\frac{D_l D_s}{D_{ls}}\right) (1+z_l) (\Delta\theta)^2 \left[ \frac{\mu-1}{\mu+1} \right] .
\end{equation}
where $\Delta\theta$ is the image separation; $\mu$ is the
magnification ratio; $z_l$ is the lens redshift; and $D_l$, $D_s$ and
$D_{ls}$ are the angular-diameter distances to the lens, to the
source, and between the lens and source, respectively. This expression
can be easily derived from more general expressions (e.g., Witt, Mao
\& Keeton 2000).

We further assume that the radio flux ratio, $f$, is a good
approximation of the magnification ratio $\mu$, even though the flux
ratio at any instant may depart from $\mu$ due to variability in the
background source, microlensing or interstellar scattering.  A
distinguishing feature of \lens~is that $f\approx 1$, which means we
may only place an upper limit on the time delay. Judging from the
values in Table~\ref{tbl:radio}, it seems likely that $\mu$ is within
5\% of unity. For $z_s = 2$ and $z_l=1$ (the largest value consistent
with the estimate in \S~\ref{subsec:nature}), this implies $h\Delta t
< 3$ days. Smaller values of $z_l$ produce smaller time delays.

Such a short time delay would be challenging to measure, because it
would require frequent sampling, and because it is poorly matched to
the typical variability time scale of quasars (months). In any case,
it may still be valuable to monitor this system to investigate
microlensing within the lens galaxy, as has been done at both optical
wavelengths (e.g.\ QSO~2237+0305; Irwin et al.\ 1989) and radio
wavelengths (B1600+434; Koopmans \& de Bruyn 2000). Here a small time
delay is an advantage, because there is no need to wait a long time to
obtain a light curve of the lagging image to compare with that of the
leading image.

As a final remark, we note that technically a singular isothermal
sphere cannot produce a magnification ratio of unity. The symmetrical
position, in which the lens is exactly aligned with the source,
produces an Einstein ring. Small misalignments produce nearly equal
doubles with high overall magnifications. If the magnification is
indeed high, one might expect more sensitive radio images to reveal
jets emerging from the quasar cores that form arcs or even an Einstein
ring. These features would provide valuable constraints on lens
models.

Real galaxies have elongated potentials, which can also produce nearly
equal doubles. The highly elongated potential of an edge-on spiral, in
particular, tends to produce magnification ratios that are close to
unity \citep{spirals}. This would be consistent with our suspicion
based on $V-I$ color (see \S~\ref{subsec:nature}) that the lens galaxy
is a spiral galaxy. In that case, lens modeling could provide
interesting constraints on spiral galaxy structure, as it has for the
edge-on spiral lens galaxy in B1600+434 \citep{b1600}.

\acknowledgments We are grateful to Tom Muxlow and Peter Thomasson for
help with the MERLIN observations, and to Jim Lovell for the ATCA
measurements. J.N.W.\ thanks the Fannie and John Hertz foundation for
supporting his graduate study, the NOAO for funding travel to Cerro
Tololo, and the Max-Planck-Institut f\"{u}r Radioastronomie for
hospitality in Bonn during part of this work.  S.L.\ acknowledges
financial support by FONDECYT grant N$^{\rm o} 3\,000\,001$ and by the
Deutsche Zentralstelle f\"ur Arbeitsvermittlung. This research was
also supported by the National Science Foundation under grants
AST-9617028 and AST-9616866.

\clearpage

\clearpage

\clearpage

\clearpage

\begin{deluxetable}{lcccccccccc}
\rotate

\tabletypesize{\scriptsize}
\tablecaption{Radio data for \lens\label{tbl:radio}}
\tablewidth{0pt}

\tablehead{
\colhead{Date}              &
\colhead{Observatory}       &
\colhead{Frequency}         &
\colhead{Beam FWHM}         &
\multicolumn{2}{c}{Flux density}      &
\colhead{RMS noise}         &
\colhead{Abs.\ flux}        &
\colhead{Flux density}        \\

\colhead{} &
\colhead{} &
\colhead{(GHz)} &
\colhead{(mas $\times$ mas, P.A.)} &
\colhead{N (mJy)} &
\colhead{S (mJy)} &
\colhead{(mJy/beam)} &
\colhead{uncertainty} &
\colhead{ratio}
}

\startdata
1998 May 18 & VLA   & 8.460  & $307 \times 167$ ($-20\fdg7$)& 14.41  & 14.97 & 0.26 & 3\% & $0.96 \pm 0.03$ \\
2000 Apr 02 & MERLIN& 4.994  & $147 \times 37$ ($15\fdg2$)  & 15.97  & 14.52 & 0.66 & 5\% & $1.10 \pm 0.07$ \\
2000 Apr 28 & VLBA  & 4.975  & $3.5 \times 1.5$ ($-0\fdg2$) & 16.37  & 15.39 & 0.22 & 5\% & $1.07 \pm 0.07$ \\
2000 Nov 01 & VLA   & 14.94  & $176 \times 108$ ($15\fdg2$) & 10.47  & 10.06 & 0.44 & 5\% & $1.04 \pm 0.06$ \\
2000 Nov 01 & VLA   & 24.46  & $122 \times 84$ ($22\fdg5$) & 8.34   & 8.07  & 0.39 &10\% & $1.03 \pm 0.09$ \\
\enddata

\tablecomments{ The most precise values for the separation between NE
and SW come from the VLBA data: $\Delta$R.A.\ $=981.47\pm 0.08$ mas,
$\Delta$Decl.\ $=552.34\pm 0.15$ mas.  All of the other radio data are
consistent with these values.  }

\end{deluxetable}

\begin{deluxetable}{lcccc}
\tabletypesize{\scriptsize}
\tablecaption{Journal of optical observations\label{tbl:journal}}
\tablewidth{0pt}

\tablehead{
\colhead{Filter} &
\colhead{Duration} &
\colhead{Seeing} &
\colhead{Airmass} \\

\colhead{} &
\colhead{(sec)}&
\colhead{(arcsec)} &
\colhead{}
}

\startdata
$I$ & 600 & 1.07 & 1.23 \\
$I$ & 600 & 1.15 & 1.20 \\
$B$ & 600 & 1.13 & 1.18 \\
$V$ & 600 & 1.00 & 1.17 \\
$R$ & 600 & 1.24 & 1.16
\enddata

\end{deluxetable}

\begin{deluxetable}{lcc}
\tabletypesize{\scriptsize}
\tablecaption{Photometric model\label{tbl:photomodel}}
\tablewidth{0pt}

\tablehead{
\colhead{Component} &
\colhead{$\Delta$R.A.} &
\colhead{$\Delta$Decl.} \\

\colhead{} &
\colhead{(mas)} &
\colhead{(mas)}
}

\startdata
NE & $490.7$  & $276.2$     \\
SW & $-490.7$ & $-276.2$  \\
G1 & $0$      & $0$   \\
G2 & $-28.5$  & $1028.7$  \\
G3 & $617.8$    & $-444.8$
\enddata

\end{deluxetable}

\begin{deluxetable}{lrrrrr}
\tabletypesize{\scriptsize}
\tablecaption{Photometry of \lens~and 5 reference stars\label{tbl:photometry}}
\tablewidth{0pt}

\tablehead{
\colhead{Object} &
\colhead{$\Delta$R.A.} &
\colhead{$\Delta$Decl.} &
\colhead{$I$} &
\colhead{$R$} &
\colhead{$V$} \\

\colhead{}   &
\colhead{(arcsec)} &
\colhead{(arcsec)} &
\colhead{(mag)} &
\colhead{(mag)} &
\colhead{(mag)}
}

\startdata
NE   & $   0.0$ & $   0.0$ & 22.328( 92) & 23.639( 91) & 24.631(230) \\
SW   & $  -1.0$ & $  -0.6$ & 22.339( 84) & 23.929(155) & $>24.6$     \\
Gal. & $  -0.5$ & $  -0.3$ & 21.891(125) & 23.113(175) & 23.830(197) \\
1    & $ 108.6$ & $  21.9$ & 19.289(  6) & 19.805(  6) & 20.432(  8) \\
2    & $  27.2$ & $ -43.6$ & 18.951(  5) & 19.738(  5) & 20.706(  7) \\
3    & $  -0.2$ & $   7.5$ & 19.912( 12) & 20.838(  9) & 21.901( 18) \\
4    & $ -16.5$ & $  78.8$ & 19.385(  7) & 20.994(  9) & 22.472( 25) \\
5    & $ -72.9$ & $ -12.5$ & 18.179(  7) & 19.002(  4) & 19.985(  5)
\enddata

\tablecomments{ The parentheses contain the statistical error in
millimagnitudes, which does not include the overall calibration error
of 0.05 magnitudes. The galaxy magnitudes are based on the sum of the
fluxes of G1, G2, and G3 in the model of Table~\ref{tbl:photomodel}.
Based on radio data, the coordinates of NE are R.A.~(J2000)$~=
20^{{\mathrm h}}04^{{\mathrm m}}07^{{\mathrm s}}.091$, Dec.~(J2000)~$=
-13\arcdeg 49\arcmin 31\farcs07$, within $0\farcs1$. }

\end{deluxetable}

\end{document}